\documentstyle[12pt]{article}

\begin{document}
\begin{center}
\large {\bf On the extension of Jackiw's scalar theory to
(2+1)-dimensional gravity}
\end{center}
\centerline{V. B. Bezerra\footnote{e-mail: valdir@fisica.ufpb.br}, 
J. Spinelly\footnote{e-mail: spinelly@fisica.ufpb.br} and 
C. Romero\footnote{e-mail: cromero@fisica.ufpb.br} } 
\begin{center}
Departamento de F\'{\i}sica, Universidade Federal da Para\'{\i}ba\\
Caixa Postal 5008, 58051-970 Jo\~{a}o Pessoa,PB,Brazil\\
\end{center}

\centerline {\bf Abstract}
\bigskip

We study some aspects of three-dimensional gravity by extending Jackiw's 
scalar theory to $(2+1)$-dimensions and find a black hole solution. We show
that in general this theory does not possess a Newtonian limit except for 
special metric configurations.


\newpage 

\section{Introduction}

Three-dimensional Einstein gravity has recently developed into an
area of active research. In this framework it was showed that\cite{Deser},
although the presence of mass can not induce curvature (locally curvature
vanishes everywhere except at the sources), it does affect the space around
the particle . The geometry around a point particle is locally flat, but 
conical in form, with a deficit angle proportional to the particle's mass. 
Thus, a system of gravitating point particle sources only affects geometry 
globally rather than locally and, as a consequence, local dynamics is 
replaced by global effects. Quantities such as the total energy-momentum and 
angular momentum for a system of point particles are defined by global 
geometric properties of the space-time surrounding the sources\cite{Deser}. 
In other words, curvature is created by sources, but only locally at their 
positions; elsewhere space-time remains flat and for this reason there can 
exist no interaction between sources. As there are no effects of gravity 
outside matter, light emitted from the surface of a star will always scape to 
infinity, and therefore, black holes do not exist in three-dimensional 
Einstein gravity. Moreover gravity does not propagate outside matter: test 
particles placed in vacuum do not experience any acceleration. On the other 
hand, Newton's gravity theory in this dimension predicts a logarithmic 
gravitational potential outside matter. As a consequence, a test particle 
placed in vacuum always accelerates in Newton's theory. For this reason 
Einstein's theory in $(2+1)$-dimensional space-time cannot reduce to 
Newton's theory by means of linearization.  

In order to overcome the problem concerning the non-existence of Newtonian
limit in this dimension some ideas have been presented, such as the 
construction of a teleparallel theory\cite{Kawai} or weakening Einstein 
equations\cite{Romero} in the same way as did Jackiw\cite{Jackiw} in his 
formulation of gravity in $(1+1)$-dimensions. In two dimensions, Einstein's 
theory does not exist because in this case the Einstein tensor 
$G_{\mu \nu}= R_{\mu \nu }-\frac 12g_{\mu \nu }R$ vanishes 
identically and the Einstein-Hilbert action  $\int d^2 x {\sqrt g}R$ is a 
surface term - it is the Euler topological invariant and does not lead to 
equations of motion.

In the middle of eighties, Jackiw and Teitelboim\cite{Teitelboim} suggested
that a way to obtain a non-trivial dynamics in $(1+1)$-dimensions is to
introduce an additional gravitational variable $\eta$ (scalar Lagrange
multiplier) and construct the following non-trivial action 
\begin{equation}
\int d^2 x {\sqrt g} \eta (R - \Lambda),  \label{1}
\end{equation}
where $R$ is the Ricci scalar and $\Lambda$ is the cosmological constant.In 
this way we can obtain a non-trivial theory of gravity in two dimensions
with field equations\cite{Jackiw,Brown}

\begin{equation}
R - \Lambda=T,  \label{2}
\end{equation}
where $T$ is the trace of the matter stress-energy tensor. 

\section{ Extending Jackiw's scalar theory}

As assumed by Romero and Dahia\cite{Romero}, we will consider that the  
eq.(\ref{2}) with $\Lambda = 0$ describes gravity in $(2+1)$-dimensions. 
Thus for source free regions we have 

\begin{equation}
R=0.  \label{3}
\end{equation}

Now, let us consider the problem of finding the motion of a test particle
under the influence of a static, circularly symmetrical matter distribution.

Differently from Romero and Dahia\cite{Romero}, who considered a conformally
flat metric that solves eq.(\ref{3}), let us choose the static, circularly
symmetrical line element given by 

\begin{equation}
ds^2=-A(r)dt^2+B(r)dr^2+r^2d\theta ^2,  \label{4}
\end{equation}
with $A$ and $B$ being functions of $r$ only. Of course, if $A$ and $B$ are 
independent, then they cannot be determined by eq.(\ref{3}) alone. To 
overcome this difficulty let us reduce the number of degrees of freedom of 
the geometry by choosing a metric tensor with only one degree of freedom. 
This can be done if we take $B=A ^{-1}$. In this case, the line element
(\ref{4}) becomes

\begin{equation}
ds^2=-Adt^2+\left( A\right) ^{-1}dr^2+r^2d\theta ^2.  \label{5}
\end{equation}

Putting eq.(\ref{5}) into (\ref{3}), leads to

\begin{equation}
\frac{d^2A}{dr^2}+\frac 2r\frac{dA}{dr}=0,  \label{6}
\end{equation}

We can imediately write down the solution of eq.(\ref{6}), which is given by

\begin{equation}
A=a+\frac br,  \label{7}
\end{equation}
where $a$ and $b$ are constants. 

Therefore, by requiring asymptotical flatness the line element in the 
present case can be written as

\begin{equation}
ds^2=-\left( 1-\frac{2\beta }r\right) dt^2+\left( 1-\frac{2\beta }r\right)
^{-1}dr^2+r^2d\theta ^2,  \label{8}
\end{equation}
where we have chosen $a=1$ and $b=-2\beta .$ 

Surely, (\ref{8}) represents the $(2+1)$-dimensional analogue of the 
Schwarzschild space-time. A simple look at the light cones structure reveals 
that the surface $r=2\beta$ acts as an event horizon. At the value $r=2\beta$ 
we have a removable coordinate singularity. On the other hand $r=0$ 
represents an essential singularity, as can be seen directly from the Riemann 
tensor scalar invariant $R_{\mu \nu \lambda \rho}R^{\mu \nu \lambda \rho}= 
\frac{24{\beta}^2}{r^{6}}$.

Therefore, the extension of Jackiw's scalar theory to $(2+1)$-dimensions
allows the existence of black hole solutions, which are not predicted by 
three-dimensional Einstein gravity, except for the case of non-vanishing
cosmological constant\cite{Mann}.
  
For the metric (\ref{8}), the geodesic equations of motion are 

\begin{equation}
\frac d{d\lambda }\left[ \left( -1+\frac{2\beta }r\right) 
\frac{dt}{d\lambda }\right] =0,  \label{11}
\end{equation}

\begin{equation}
\frac d{d\lambda }\left[ r^2\frac{d\theta }{d\lambda }\right] =0
\label{12}
\end{equation}

and

\begin{equation}
-\left( 1-\frac{2\beta }r\right) \left( \frac{dt}{d\lambda }\right)
^2+\left( 1-\frac{2\beta }r\right) ^{-1}\left( \frac{dr}{d\lambda }\right)
^2+r^2\left( \frac{d\theta }{d\lambda }\right) ^2=-\varepsilon .
\label{13}
\end{equation}
where $\varepsilon$ is a constant that takes the values $-1, 0, 1$, for
space-like, null and time-like curves, respectively. 

Integrating (\ref{11}) and (\ref{12}) yields 

\begin{equation}
\left( 1-\frac{2\beta }r\right) \frac{dt}{d\lambda } =E  \label{14}
\end{equation}

and

\begin{equation}
r^2\frac{d\theta }{d\lambda }=L,  \label{15}
\end{equation}
where $E$ and $L$ are integration constants.

For $\varepsilon=1$ and $\lambda=\tau$ ($\tau$ being the proper time as 
measured by a particle following geodesic motion) it follows from 
eqs.(\ref{13}), (\ref{14}) and (\ref{15}) that

\begin{equation}
\frac 12\left( \frac{dr}{d\lambda }\right) ^2+V\left( r\right) =\frac{E^2}2.
\label{16}
\end{equation}

The above equation might be loosely interpreted as the equivalent of energy
conservation in Newtonian gravity with $V\left( r\right)$ playing the
role of an "effective potential" given by

\begin{equation}
V\left( r\right) =\frac \varepsilon 2-\frac{\beta \varepsilon }r+\frac{L^2}{%
2r^2}-\frac{\beta L^2}{r^3}.  \label{17}
\end{equation}

By the same token $L$ would be looked upon as the angular momentum of the 
particle per unit mass.

The critical points of the "effective potential" (\ref{17}) can be obtained 
by solving

\begin{equation}
 \frac {\beta \varepsilon}{r^2}-\frac{L^2}{r^3} +%
\frac{3\beta L^2}{r^4}=0.  \label{18}
\end{equation}

For null geodesics ($\varepsilon=0$) it is easily seen from (\ref{16}) and
(\ref{18}) that photons can follow the circular orbits $r=3 \beta$ provided
that $E^{2}=\frac{L^2}{27 \beta^2}$. These orbits are highly unstable as
$r= 3 \beta$ is an isolated maximum for $V(r)$. If $E^{2}>\frac{L^2}{27 
\beta^2}$, then either the photon will fall towards the singularity
($r<3 \beta$) or it will receed into infinity($r> 3 \beta$).

For massive particles ($\varepsilon=1$) one or two circular orbits with radii 
$r_{\pm }=\frac{L^2\pm \left( L^4-12\beta ^2L^2\right) ^{\frac 12}}{2\beta }$
are possible according to $L^2=12\beta ^2,$ or $L^2>12\beta ^2,$ respectively.
If $L^2<12\beta ^2,$ then it is not hard to see that the particle is forced to
move towards the singularity at $r=0$.

\section{ Newtonian limit of Jackiw's scalar theory}

Given that Einstein's $(2+1)$-dimensional gravity exhibits such a drastic 
departure from the corresponding Newtonian gravity, one would wonder whether
or not the present extension of Jackiw's scalar gravity does have a Newtonian 
limit. This question is better examined if one considers Galilean coordinates, 
the ones in which

\begin{equation}
g_{\mu \nu}=\eta_{\mu \nu}+h_{\mu \nu}, \label{19}
\end{equation}
with $\eta_{\mu \nu}=diag\left( -1, 1, 1, 1 \right)$ and $|h_{\mu \nu}|<<1$. 
It is a well-known fact that for any metric theory of gravity small 
departures from Minkowski flat space-time lead to geodesic equation which, 
for a non-relativistic particle, perfectly mimic Newton's equation of motion 
in a classical gravitational field, as long as the metric tensor is time-
independent. On the other hand one expects that with the same kind of 
approximation the field equations should reduce to Poisson's equation for 
the classical field. When these two conditions are consistently fulfilled, 
then one would say that the theory has a Newtonian limit.

Let us first briefly recall how the effect of a gravitational field of force 
can be obtained by using the geodesic equations of motions under the 
circumstances described above. Thus, consider the line element of a nearly 
Minkowskian metric tensor given by

\begin{equation}
ds^2= -\left(dx^{0} \right)^2+\left(dx^{1} \right)^2+\left(dx^{2} \right)^2+
h_{\mu \nu}dx^{\mu}dx^{\nu}, \label{20}
\end{equation}
with $x^{0}=ct$, $\mu, \nu=0,1,2$. If the geodesic curve is parametrized by 
the
coordinate time $t$, then we have 

\begin{equation}
\left(\frac{ds}{dt} \right)^2=-c^2\left( 1-\beta^2-\frac{h_{\mu \nu}}{c^2}
\frac{dx^{\mu}}{dt}\frac{dx^{\nu}}{dt}\right), \label{21}
\end{equation}
where $\beta=\frac{v}{c}$ and $v$ denotes the velocity of the particle along 
the geodesic. For non-relativistic motion $\beta$ is small and, in our 
approximation, only first-order terms in $\beta$ and $h_{\mu \nu}$ will be 
retained. Thus, to first order in $h_{\mu \nu}$ and $\beta$, (\ref{21}) becomes

\begin{equation}
\left(\frac{ds}{dt} \right)^2 \simeq c^2 \left( h_{00}-1\right). \label{22}
\end{equation}
Now, let us consider the geodesic equations

\begin{equation}
\frac{d^{2}x^{\mu}}{ds^{2}}+\Gamma^{\mu}_{\alpha \beta}\frac{dx^{\alpha}}
{ds}\frac{dx^{\beta}}{ds}=0.   \label{23}
\end{equation}
Again, keeping only first-order terms in $h_{\mu \nu}$ and $\beta$, it can 
be easily verified that (\ref{23}) becomes

\begin{equation}
\frac{d^{2}x^{\mu}}{ds^{2}}+\Gamma^{\mu}_{00}\left(\frac{dx^{0}}{ds} 
\right)^{2}=0. \label{24}
\end{equation}
On the other hand, in this approximation, we have 

\begin{equation}
\Gamma^{i}_{00}=\frac{1}{2}\frac{\partial h_{00}}{\partial x^{i}}, \label{25}
\end{equation} 
for $i=1,2$. From (\ref{24}) and (\ref{25}) one gets

\begin{equation}
\frac{d^{2}x^{i}}{dt^{2}}=\frac{\partial}{\partial x_{i}}\left(-\frac{h_{00}}
{c^2} \right), \label{26} 
\end{equation}
which looks like Newton's equation of motion for a particle in a classical 
gravitational field provided that we identify the scalar gravitational field
as being

\begin{equation}
\phi=\frac{c^{2}}{2}h_{00}, \label{27}
\end{equation} 
and requiring that $h_{00}$ and $\phi$ vanish at infinity. At this point let 
us linearize the field equation of Jackiw's gravity theory by assuming the 
weak-field approximation (\ref{19}). If matter is present Jackiw's field 
equation is given by

\begin{equation}
R=kT, \label{28}
\end{equation}
where $T={T^{\mu}}_{\mu}$ denotes the trace of the energy-momentum 
$T_{\mu \nu}$ and $k$ is constant. In the non-relativistic regime 
$\frac{|T_{ij}|}{T_{00}}<<1$, i.e., all stress are small compared to the 
density of energy $T_{00}=\rho$. Thus, we have $T \simeq \rho$. It remains 
to find the expression for the linearized $R$. This can be done simply by 
noting that to first order in $h_{\mu \nu}$ we have 

\begin{equation}
R_{\mu \nu}= -\frac{1}{2}\left( \Box^{2}h_{\mu \nu}-\frac{\partial^{2}}
{\partial x^{\lambda} \partial x^{\mu}}{h^{\lambda}}_{\nu}- 
\frac{\partial^{2}}{\partial x^{\lambda} \partial x^{\nu}}{h^{\lambda}}_
{\mu}+\frac{\partial^{2}}{\partial x^{\mu} \partial x^{\nu}}h \right), 
\label{29}
\end{equation}  
where $h={h^{\mu}}_{\mu}$. Recalling that in this approximation indices are 
lowered and raised with the Minkowski metric tensor, from (\ref{11}) one 
readily obtains

\begin{equation}
R= -\Box^{2}h+\frac{\partial^{2}}{\partial x^{\lambda} \partial x^{\nu}}h^
{\lambda \nu}. \label{30}
\end{equation}

This equation can be simplified if we choose to work in the so-called 
harmonic coordinate system, that is, the one for which 
$ \frac{\partial h_ {\nu}^{\lambda}}{\partial x^ {\lambda}}=\frac{1}{2} 
\frac{\partial h}{\partial x^{\nu}}$.

Finally, if one assumes that the metric is not time-dependent, Jackiw field 
equation (\ref{28}) turns into

\begin{equation}
\nabla^{2}\phi={kc^{2}\rho}-\frac{c^{2}}{2}\nabla^{2}
(h_{11}+h_{22}), \label{31}
\end{equation}
where use has been made of the equation (\ref{27}). Given that the second 
term of the right-hand side of equation above does not vanish in general, 
(\ref{31}) is not identical to Poisson's equation for the classical 
gravitational field. It is worth mentioning, however, that in the static
conformally flat case $h_{\mu \nu}= \epsilon \eta_{\mu \nu}$, hence

\begin{equation}
R = \frac{3}{2} \nabla^{2} h_{00}, \label{32}
\end{equation}
and a Newtonian limit can be defined, a result which was previously obtained
by Cornish and Frenkel\cite{Cornish}. Therefore, we conclude that the 
extension of Jackiw's scalar theory to (2+1)-dimensional gravity does not 
lead to a theory with proper Newtonian limit, except for a few special metric 
configurations. 

\noindent
{\bf Acknowledgment}\\
\noindent
This work was partially supported by CNPq and CAPES. We are indebted to
F. Dahia for enlightening discussions.

\end{document}